\documentclass[a4paper]{article}

\usepackage{graphicx}
\usepackage{amsmath,amssymb}
\usepackage[all]{xy}
\newcommand{\email}[1]{\texttt{#1}}

\newcommand{\Den}{{\cal D}}

\newcommand{\EUV}{{\epsilon}}
\newcommand{\IE}{{i \varepsilon}}

\newcommand{\SetReal}{\mathbb{R}}

\newcommand{\Li}{\operatorname{Li}}

\newcommand{\Slash}[1]{{\kern-0.8ex}\not{\kern-0.4ex #1}}

\newcommand{\FD}[4]{F_D\left(#1,#2;#3;#4\right)}    

\newcommand{\Ord}[1]{{\cal O}(#1)}

\newcommand{\Comment}[1]{}

\begin{document}
\title{Numerical calculation of one-loop integration 
       with hypergeometric functions%
\footnote{
  Talk given at 3rd Computational Particle Physics Workshop -- CPP2010,
                  September 23-25, 2010,
                  KEK Japan}
}

\author{Toshiaki Kaneko%
   \footnote{E-mail: \email{toshiaki.kaneko@kek.jp}}
         \\
         High Energy Accelerator Research Organization (KEK)\\,
             Computing Research Center, \\
         1-1 Oho, Tsukuba, Ibaraki 305-0801 Japan
}

\date{~}
\vspace{2em}

\maketitle
\begin{abstract}{
One-loop two-, three- and four-point scalar functions
are analytically integrated directly such that they are expressed in terms of
Lauricella's hypergeometric function $F_D$.
For two- and three-point functions, exact expressions are obtained with
arbitrary combination of kinematic and mass parameters in arbitrary 
space-time dimension.
Four-point function is expressed in terms of $F_D$ up to the finite
part in the expansion around 4-dimensional space-time with
arbitrary combination of kinematic and mass parameters.
Since the location of the possible singularities of $F_D$ is
known, information about the stabilities in the numerical
calculation is obtained.
We have developed a numerical library calculating $F_D$
around 4-dimensional space-time.
The numerical values for IR divergent cases of four-point functions
in massless QCD are calculated and agreed with \texttt{golem95} package.
}
\end{abstract}
\vspace{2em}

\section{Introduction}

One-loop calculation in perturbative field theory is a well-established
and theoretically clear method.
However, highly accurate numerical calculations of one-loop amplitudes
are not a trivial problem.
There appear many kinematic parameters including particle masses.
It is not easy to keep numerical accuracy for every points in
the multi-dimensional parameter space.
Since the numerical behavior of a function is related to
its analytic properties,
it is important to pursue stable analytical expressions for
numerical calculations.

Feynman amplitudes are expected to be a kind of hypergeometric
function%
\cite{phys:Regge-1968}.
One-loop integrals are explicitly expressed in terms of
hypergeometric functions using several different methods:
Mellin-Bares transformation,
algebraic relations and
power series expansion %
\cite{phys:Davydychev,
      phys:Fleischer-Jegerlehner-Tarasov,
      phys:Duplancic-Nizic-2001,
      phys:Kurihara-2006}.
See also \cite{phys:KKWY} for the references.
In this article, we show that one-loop two-, three- and
four-point functions are directly integrated with
Gauss' function $F$, Appell's function $F_1$ and 
Lauricella's $F_D$%
\cite{math:Erdeley-Magnus-ETAL-1953}.
Among these function, $F_D$ includes other functions as special cases.
Since the location of possible singularities of $F_D$ is known%
\cite{math:Iwasaki-etal-1991},
we can point out dangerous combinations of parameters 
and to get information about
where and how numerical cancellation may occur.
For two- or three-point functions, we show the integrals are exactly
expressed in terms of $F_D$ for any values of kinematic
parameters in any space-time dimensions.
For four-point case, we could not integrate exactly except some
special cases.
However, around 4-dimensional space-time,
it is expressed in terms of $F_D$ up to the finite order.

We have developed a program package which calculate $F_D$
for the necessary combinations of parameters for IR
divergent case in massless QCD.
We have compared our numerical result with \texttt{golem95} package%
\cite{phys:Golem95}.

\section{Two-point point function}

Let's first consider two-point scalar function as a simple example of the usage
of hypergeometric function.
Two-point function is defined by:
\begin{equation} \label{twopnt1}
\begin{split}
I_2^{(\alpha)} &= 
      \int_{0}^{\infty} dx_1 \, \int_{0}^{\infty} dx_2 \; 
      \delta(1-x_1-x_2) \; \Den^{\alpha}
,\\
\Den &=
         - p^2 x_1 x_2 + m_1^2 x_1 + m_2^2 x_2 - \IE
,
\end{split}
\end{equation}
where, $\alpha$ is a number depending on the dimension of the
space-time.
It is noted that calculation with arbitrary mass parameters is
useful even for massless cases, since tensor integrations are
obtained by differentiating scaler one in terms of mass parameters.

Integrating once with the $\delta$ function, we obtain
\begin{equation} \label{twopnt2}
\begin{split}
I_2^{(\alpha)} &= 
     (m_2^2)^{\alpha}
     \int_0^1 
         \left(1 - \frac{x}{\gamma^{+}}\right)^\alpha
         \left(1 - \frac{x}{\gamma^{-}}\right)^\alpha
         \; dx
,\\
\Den &= p^2 x^2 + ( - p^2 + m_1^2 - m_2^2 ) x + m_2^2 
      = m_2^2
         \left(1 - \frac{x}{\gamma^{+}}\right) 
         \left(1 - \frac{x}{\gamma^{-}}\right)
,\\
\gamma^\pm &= \frac{p^2-m_1^2+m_2^2 \pm \sqrt{D}}{2 p^2}
,\qquad
D = (-p^2+m_1^2+m_2^2)^2 - 4 m_1^2 m_2^2
.
\end{split}
\end{equation}
This integral is nothing but a special case of Appell's $F_1$, 
whose integral form is:
\begin{equation}
  F_1(a, b, b'; c; y, z)
=
  \frac{\Gamma(c)}{\Gamma(a)\Gamma(c-a)}
  \int_0^{1} x^{a-1} (1-x)^{c-a-1}
             (1-yx)^{-b} (1-zx)^{-b'} \; dx 
.
\end{equation}
Thus we have
\begin{equation} \label{twopnt3}
I_2^{(\alpha)}
  =  (m_2^2)^{\alpha}
      F_1(1, - \alpha, - \alpha; 2;
         \frac{1}{\gamma^{+}}, \frac{1}{\gamma^{-}})
  =  (m_1^2)^{\alpha}
      F_1(1, - \alpha, - \alpha; 2;
         \frac{1}{1-\gamma^{+}}, \frac{1}{1-\gamma^{-}})
.
\end{equation}
The last equality is obtained by changing integration variable
form $x$ to $y = 1-x$.
This equality is considered as an identity of $F_1$.

Function $F_1$ reduces to Gauss' hypergeometric
function $F$ for the case of $b' = 0$ and
is a special case of more general function,
called Lauricella's $F_D$, whose integral form is:
\begin{equation}
\begin{split} \label{fd}
F_D(a,&\,b_1,\cdots,b_n;\,c;\,z_1,\cdots,z_n) =
  \frac{\Gamma(c)}{\Gamma(a)\Gamma(c-a)}
  \int_0^1 x^{a-1} (1-x)^{c-a-1}
    \prod_{i=1}^n (1-z_i x)^{-b_i} dx
.
\end{split}
\end{equation}
This function $F_D$ will be used for vertex and box integration.
The location of possible singularities of $F_D$
is limited to
$z_i = 0, 1, \infty$ and $z_j = z_k\;(j \neq k)$
and $F_D$ is smooth except these special cases%
\cite{math:Iwasaki-etal-1991}.
The power series expansion and differential equation of this function 
is known.
Many of identities of $F$ can be generalized to $F_D$.
We especially use the following identity, which can easily be
confirmed from Eq.(\ref{fd}):
\begin{equation} \label{fddiff}
\begin{split}
z^{p-1} & (1-z)^{q-1} \prod_{i=1}^{n-1} (1 - x_i z)^{-b_i}
=
\frac{d~}{d z} \frac{z^p}{p}
  \FD{p}{(b_i), 1-q}{p+1}{(x_i z), z}
.
\end{split}
\end{equation}
This identity implies that any product of linear factors with
arbitrary power is integrated by $F_D$. 

Now going back to Eq.(\ref{twopnt3}).
Two-point function $I_2^{(\alpha)}$ may be singular 
when $\gamma^{\pm} = 0, 1, \infty$
or $\gamma^+ = \gamma^-$.
These cases correspond to massless particles, $p^2 = 0$, and on the threshold.
Let us examine the case of $m_2^2 = 0$ and 
$m_1^2, p^2 \neq 0$ 
for an example, where
$\gamma^- = 0$ and $\gamma^+ = (p^2-m_1^2)/p^2$.
Although Eq.(\ref{twopnt3}) is not well-defined for the limit of
$m_2 \rightarrow 0$,
we can use alternative representation obtained with another
identity of $F_1$:
\begin{equation} \label{twopnt4}
\begin{split}
I_2^{(\alpha)} &=
      \frac{\gamma^{-}}{\alpha+1}
      (m_2^2)^\alpha
      F(\alpha+1, -\alpha; \alpha+2;
        \frac{\gamma^{-}}{\gamma^{-}-\gamma^{+}})
\\   &
    + \frac{1-\gamma^{-}}{\alpha+1}
      (m_1^2)^\alpha
      F(\alpha+1, -\alpha; \alpha+2;
        \frac{1-\gamma^{-}}{\gamma^{+}-\gamma^{-}})
.
\end{split}
\end{equation}
This representation is regular for the limit of $m_2 \rightarrow 0$
under the condition of $\Re \alpha > 0$.
Then expanding around 4-dimensional space-time 
( $\alpha = - \epsilon \rightarrow +0$ ), 
$F_1$ is reduced to $F$ and then to $\log$ and $\Li_2$.
Two limiting processes $m_2 \rightarrow 0$ and $\alpha \rightarrow +0$
do not commute.
This property corresponds to the fact that the analytic result
in 4-dimension does not reduces to the massless one by taking the
simple limit $m_2 \rightarrow 0$.

This example shows that Eq.(\ref{twopnt3})
can be used as a unified representation for both
massive and massless cases.
When values of mass parameters are specified,
we select a suitable representation with appropriate limit.
We want to select it not on a notebook but in a numerical library at
the time of numerical calculations.
With this library,
the main program will be general to cover various cases of
parameters.
The problem of numerical instability will be confined
into the numerical calculation method of $F_D$.

\section{Three-point function}

Three-point function is defined by:
\begin{equation}
I_3^{(\alpha)} =
   \int_{x_1, x_2 > 0, x_1 + x_2 < 1} dx_1 \; dx_2 \; \Den^\alpha
.
\end{equation}
where, $\Den$ is a quadratic form of $x_1$ and $x_2$.
We apply the projective transformation as shown by 
Ref. \cite{phys:tHooft-Veltman-1979}.
The quadratic term of $x_2$ is eliminated by changing variables
$(x_1, x_2) \rightarrow (x_2, z = x_1 + r x_2)$
with adjusting of the value of $r$.
Since $\Den$ is now linear in $x_2$, integration is trivial for $x_2$.
The resulting integration becomes the form:
\begin{equation}
I_3^{(\alpha)} \propto
\int \frac{\Den^{\alpha+1}}{a z + b} \; d z
.
\end{equation}
As $\Den$ is expressed as a product of linear factors of $z$,
we obtain:
\begin{equation}
I_3^{(\alpha)} \propto
   \int \frac{1}{a z + b} 
         \left(1 - \frac{z}{\gamma^+}\right)^{\alpha+1}
         \left(1 - \frac{z}{\gamma^-}\right)^{\alpha+1}
   \; d z
.
\end{equation}
Using Eq.(\ref{fddiff}), this is immediately integrated by $F_D$.
The integration domain becomes slightly complicated after the
projective transformation.
It is handled systematically with exterior derivative and
Stokes' theorem.
The result takes the following form:
\begin{equation} 
I_3^{(\alpha)}  =
   \frac{1}{\alpha+1} 
   \sum_{k=0}^{2} 
   \frac{\Den_k^{\alpha+1}}{a}
    \frac{d_{k,1}}{d_{k,0}}
    F_D(1, 1, -\alpha-1, -\alpha-1; 2;
         - \frac{d_{k,1}}{d_{k,0}},
           \frac{1}{\gamma_k^{+}},
           \frac{1}{\gamma_k^{-}})
\end{equation} 
where, $\Den_k$ is the value of $\Den$ at the corner of the
integration domain and $d_{k,j}$ is brought from the parameterization 
of the boundary after the projective transformation.
For the limit of 4-dimensional space-time,
the above expression reduces to the usual analytic representation
as $F_D$ reduces to 
$F_D \rightarrow F_1 \rightarrow F \rightarrow \log$ 
and poly-logarithmic functions.

\section{Four-point function}

Four-point function is written by:
\begin{equation}
I_4^{(\alpha)} = \int_{\SetReal_{\geq 0}^4} d^4 x \;
             \delta\left(1 - \sum_{j=1}^4 x_j\right) \; \Den^\alpha
.
\end{equation}
where $\Den$ is a homogeneous quadratic form of $x_j$.
After using $\delta$ function, there left three integration variables.

We apply projective transformations twice.
After the first transformation, we integrate once
using the following identity:
\begin{equation}
\int \Den^{\alpha} d y = 
  \frac{1}{\alpha + 1} \frac{\Den^{\alpha+1}}{\partial_y \Den}
,
\end{equation}
where $\Den$ is quadratic in terms of remaining two variables,
while $\partial_y \Den$ is linear.

The projective transformation is applied once more.
$\Den$ becomes a linear function of a new variable $z$.
It is possible to select the variable $z$ 
by shifting and rescaling 
such that $z \propto \Den$ and 
$1-z \propto \partial_y \Den$.
Integral is calculated with the following formula 
(special case of Eq. (\ref{fddiff})):
\begin{equation}
b \, \frac{z^{b-1}}{1-z}
= \frac{d~}{d z} z^{b} F(1, b; b+1; z) 
,
\end{equation}
where $F$ is Gauss' hypergeometric function.
After the second integration we obtain:
\begin{align}
I_4^{(\alpha)} 
 &= 
    \sum_{k=1}^3 \;
    \sum_{\ell=1, \ell\neq k}^4 \;
    \xi_k^{(4)} 
    \xi_\ell^{(k)} 
   \int_{L_{k\ell}} [ g_k + h_k(e_k) ]\; d y_{k\ell} ,
\\
g_k &= \frac{1}{(\alpha+1)(\alpha+2)}
   \frac{e_k^{\alpha+1}}{d_k^{\alpha+2}}
   \left(\frac{d_k \Den_k}{e_k}\right)^{\alpha+2}
   F(1,\alpha+2,\alpha+3; \frac{d_k\Den_k}{e_k})
,
\end{align}
where
$\Den_k$ and $e_k$ are quadratic form of integration variable $y_{kl}$,
$d_k$ and $\xi_\ell^{(k)}$ are brought by projective transformation,
and
$L_{k\ell}$ is a line segment of the last integration.
Function $h_k$ of $e_k$ is arbitrary and is produced as 
an integration constant.

In order to handle $F$ in the integrand, we use partial integration
method.
Using recursion relation of $F$, $g_k$ is expressed by:
\begin{equation}
\begin{split}
g_k &= 
   \frac{1}{(\alpha+1)(\alpha+2)}
   \frac{e_k^{\alpha+1}}{d_k^{\alpha+2}}
   \left(\frac{d_k \Den_k}{e_k}\right)^{\alpha+2}
\\ &\quad
 + \frac{1}{(\alpha+1)(\alpha+3)}
   \frac{e_k^{\alpha+1}}{d_k^{\alpha+2}}
   \left(\frac{d_k \Den_k}{e_k}\right)^{\alpha+3}
   F(1,\alpha+3,\alpha+4; \frac{d_k\Den_k}{e_k})
\end{split}
\end{equation}
Factor $e_k^{\alpha+1}$ is integrated by the following relations:
\begin{align}
e_k(x) &= \tilde{e}_k \, (w_5 - x) (w_6 - x)
,\qquad
e_k^{\alpha+1}(x)  
  = \frac{d f(x)}{d x}
,\\ \nonumber
f(x) &= 
   \frac{1}{\alpha+2}
     \frac{e_k^{\alpha+2}(x)}{\tilde{e}_k(w_6-w_5)}
   \Bigl[
   1
 - 2 F(-\alpha-2, \alpha+2, \alpha+3; \frac{w_5-x}{w_6-x})
   \Bigr]
.
\end{align}
After the partial integration, we obtain:
\begin{equation}
\begin{split}
I_4^{(\alpha)} 
  &= 
    \sum_{k=1, k\neq m}^4 \;
    \sum_{\ell=1, \ell\neq k}^4 \;
    \xi_k^{(m)} 
    \xi_\ell^{(k)} 
   [J_1 + J_2 + J_3] ,
\\
  J_1 &=
   \frac{1}{(\alpha+1)(\alpha+2)}
   \int_0^1
   \frac{{\Den_k}^{\alpha+2}}{e_k}
   \; dx ,
\\
  J_2 &=
   \frac{1}{(\alpha+1)(\alpha+3)}
   \frac{1}{d_k^{\alpha+2}}
   \Bigl[
   f(x)
   \left(\frac{d_k \Den_k}{e_k}\right)^{\alpha+3}
   F(1,\alpha+3,\alpha+4; \frac{d_k\Den_k}{e_k})
   \Bigr]_{x=0}^{1} ,
\\
  J_3 &=
 - \frac{1}{\alpha+1}
   \frac{1}{d_k^{\alpha+2}}
   \int_0^1
   f(x)
   \left(\frac{d_k \Den_k}{e_k}\right)^{\alpha+2}
   \frac{d~}{d x} \log \left[\frac{e_k - d_k \Den_k}{e_k}\right]
   \; d x .
\end{split}
\end{equation}
Here we have used the following identity:
\begin{equation}
\frac{d~}{d x} R(x)^{a+3} F(1,a+3,a+4;R(x))
= - (a+3) R(x)^{a+2} \frac{d~}{d x} \log [1-R(x)]
.
\end{equation}
$J_1$ is integrable with $F_D$, since the integrand is
expressed by a product of power of linear factors.
$J_2$ is a product of $F$.
The problem left is function $f$ in $J_3$.

Investigating the limit to the 4-dimensional space-time
$\alpha = - 2 - \EUV \rightarrow - 2$,
one will confirm that integration in $J_3$ does not produce
new poles of $1/\EUV$.
So we can expand in terms of $\EUV$ in the integrand.
\begin{gather}
F(\EUV, -\EUV, 1-\EUV; z)
 = 1 + \Ord{\EUV^2}
.
\\
J_3 =
   \frac{1}{\EUV(1+\EUV)}
   \frac{1}{\tilde{e}_k(w_6-w_5)}
   \int_0^1
   {\Den_k}^{\alpha+2}
   \frac{d~}{d x} \log \left[\frac{e_k - d_k \Den_k}{e_k}\right]
   \; dx
   + \Ord{\EUV}
.
\end{gather}
Since factor $d \log / d x$ is expressed by a sum of
inverse of linear term of $x$, 
$J_3$ is expressed by $F_D$ up to the finite order.

\section{Sample numerical calculation}

Since $F$, $F_1$ and $F_D$ have many parameters and variables,
it is hard to construct numerical package
to calculate for all cases.
However,
we need numerical values only for some special combination of parameters
for our purpose.

We have tried sample numerical calculations of one-loop box 
tensor integration
for the cases of Ref. \cite{phys:Duplancic-Nizic-2001}.
That is:
\begin{itemize}
\item
All particles are massless.

\item
At least one external particle is on-shell ($p_1^2 = 0$).

\item 
Calculate 
up to $\Ord{\EUV^0}$.

\item 
Calculate tensor integrations up to 
$\text{ rank} = 4$.

\end{itemize}
There appear IR divergences, which are represented by the poles of
$1/\EUV$.

For these cases, integration becomes simpler by using variable
transformation as described in
\cite{phys:Duplancic-Nizic-2001}.
We have obtained an exact analytic representation with $F_D$
for the cases of 
  4 or 3 on-shell particles
and
  ``easy case'' of 2 on-shell  particles
(diagonal external particles of the box diagram are on-shell).
However, for 
  ``hard case'' of 2 on-shell particles
(two adjacent external particles of the box diagram are on-shell)
and 
  3 on-shell case, 
we need to expand in terms of $\EUV$.

It is noted that representations of $n$-point functions are not
necessarily numerically stable when they are written with poly-logarithmic
functions.
For example, there appears $F(1, m-\EUV, m+1-\EUV; z)\;(m \geq 1)$
in the tensor integrations, which is regular around $z \sim 0$.
When this function is expanded in terms of $\EUV$ using
identities of $F$, the following combination of terms appears:
\begin{gather*}
   \frac{\EUV^k}{z^m}
   \left[ \Li_{k+1}(z) - \sum_{j=1}^{m-1} \frac{z^j}{j^{k+1}} \right]
.
\end{gather*}
When $\Li_{k+1}$ is expanded around $z \sim 0$,
the first $m-1$ terms of the power series
cancel out with the second term in the brackets.
Factor $z^m$ is factored out from the resulting terms in the
bracket,
and cancels with the denominator.
If these terms are scattered into a long expression,
it is not an easy problem to control these numerical
cancellations and the singular behavior of the denominator.
However, when we keep the original form of
$F(1,m-\EUV,m+1-\EUV;z)$, the problem is immediately solved;
a simple power series calculation of $F$ around $z = 0$
produces stable result.
This problem is caused by the expansion made in order to express these
functions in terms of $\Li_k$.

We have developed a numerical library of hypergeometric function
$F_D$ for
necessary combination of parameters for our sample calculations.
Our library is designed in the following way:
\begin{enumerate}
\item 
Entry points of subroutines are $F_D$ or $F$.

\item 
These subroutines return an array of coefficients
of $1/\EUV^2, 1/\EUV, 1, \EUV, \cdots$ 
up to necessary order.

\item 
Inside of the subroutines, appropriate identities or calculation methods
are selected in looking
at the values of parameters and variables.

\end{enumerate}

Two programs are prepared for the numerical calculations of
tensor integrations:
\begin{itemize}
\item
``program-1'':
calculation with numerical library of $F$, $F_1$ and $F_D$.

\item
``program-2'':
calculation with the following numerical integration:
\begin{itemize}
\item 
The first two integrations are calculated analytically.

\item 
Coefficients of \(1/\EUV^2, 1/\EUV^1, 1/\EUV^0\) are
extracted and expressed by one-dimensional integrations.

\item
The last integration is calculated numerically (Romberg method).
\end{itemize}
\end{itemize}

We have compared the numerical results
among program-1,  program-2
and \texttt{golem95} package up to $\text{ rank} = 4$
at 7560 different values of the parameters:
\begin{gather*}
p_1^2 = 0
, \qquad
p_2^2 = 0, \pm 50
, \qquad
p_3^2 = 0, \pm 55
, \qquad
p_4^2 = 0, \pm 60
,
\\
s = \pm 200
, \qquad
t = \pm 123
,
\\
n_i = 0, 1, 2, 3, 4
, \qquad 
\sum_i n_i \leq 4
\qquad \text{(rank of tensor integration)}
\end{gather*}
The results of the maximal differences among methods are shown 
by Table \ref{tab:num}.
It shows that the accuracy of the library seems similar to
\texttt{golem95} package.

\begin{table}[!h]
\begin{center}
\begin{tabular}{ccl}
\hline
\multicolumn{2}{c}{calculation method} & maximal difference \\
\hline
program-1(d) &  program-2(d)  & \(7.65 \times 10^{-7}\) \\
program-1(d) &  \texttt{golem95} (d)      & \(9.13 \times 10^{-10}\) \\
program-1(d) &  program-1(q)  & \(3.98 \times 10^{-10}\) \\
\texttt{golem95}(d) & \texttt{golem95}(q) & \(5.17 \times 10^{-10}\) \\
program-1(q) &  \texttt{golem95} (q)      & \(1.38 \times 10^{-18}\) \\
\hline
\end{tabular}
\end{center}
\caption{Maximal differences among the calculation method.
Differences are measured by the distance on the complex plane.
(d) and (q) stand for double and quadruple recision respectively.
}
\label{tab:num}
\end{table}

\section{Summary}

Two- and three-point functions are expressed in terms of $F_D$,
exactly for any combination of physical parameters in
any space-time dimensions.
Four-point functions are expressed with $F_D$,
up to $\Ord{\EUV^0}$ for any combination of physical parameters.
A program library of $F$ and $F_D$ is developed applicable 
for sample numerical calculations for massless QCD box with IR
divergences.
The results agree with \texttt{golem95} package.
Four-point function seems not to be integrated with $F_D$.
In order to express general four-point function, 
more general hypergeometric functions will be needed
as described in %
\cite{phys:Davydychev,
      phys:Fleischer-Jegerlehner-Tarasov}
and \cite{math:GKZ}
for more general cases.

\vspace{1em}
\noindent
\textbf{Acknowledgments}

The author wish to express his thanks to the members
of Minami-tateya group.
He is especially indebted to Y.~Kurihara for stimulus
and useful discussions.
This work is supported by
Ministry of Education, Science, and Culture, Japan
under Grant-in-Aid No.20340063 and No.21540286.


\begin{thebibliography}{99}

\bibitem{phys:Regge-1968}
Tullio Regge,
{\em in \textsl{Battelle Rencontres, 1967 Lectures in Mathematics and
  Physics}, ed. C.~M.~DeWitt and J.~A.~Wheeler}, (1968) 433--458.

\bibitem{phys:Davydychev}
A.~I.~Davydychev,
J.~Math.~Phys. \textbf{32} (1991) 1052--1060;

A.~I.~Davydychev,
J.~Math.~Phys. \textbf{33} (1992) 385--369.

\bibitem{phys:Fleischer-Jegerlehner-Tarasov}
O.~V.~Tarasov,
Phys. Rev. D \textbf{54} (1996) 6479--6490;



\bibitem{phys:Duplancic-Nizic-2001}
G.~Duplan\v{c}i\'{c} and B.~Ni\v{z}i\'{c},
{\em Eur. Phys. J.} \textbf{C 20} (2001) 347--370.

\bibitem{phys:Kurihara-2006}
Y.~Kurihara,
{\em Eur. Phys. J.} \textbf{C 45} (2006) 427--444.

\bibitem{phys:KKWY}
M.~Yu.~Kalmikov, B.~A.~Kniehl, B.~F.~W.~Ward and S.~A.~Yost,
\textsl{Hypergeometric functions, their $\epsilon$ expansions
and Feynman diagrams},
arXiv:0810.3238.

\bibitem{math:Erdeley-Magnus-ETAL-1953}
A.~Erd\'eley, W.~Magnus, F.~Oberhettinger, and F.G. Tricomi,
\textsl{Higher transcendental functions, I--III},
Bateman Manuscript Project (1953--1955).

\bibitem{math:Iwasaki-etal-1991}
K.~Iwasaki, H.~Kimura, S.~Shimomura, and M.~Yoshida,
\textsl{From Gauss to Painlev\'e : a modern theory of special
  functions},
Vieweg (1991).

\bibitem{phys:tHooft-Veltman-1979}
G.~'tHooft and M.~Veltman,
{\em Nucl. Phys.} \textbf{B 153} (1979) 365--401.

\bibitem{phys:Golem95}
T.~Binoth et~al.
{\em Comput. Phys. Commun.} \textbf{180} (2009) 2317.




\bibitem{math:GKZ}
I.~M.~Gel'fand, A.~V.~Zelevinsky, M.~M.~Kapranov,
{\em Funk. Anal. Appl.} \textbf{23} (1989) 94--106;

I.~M.~Gel'fand, M.~M.~Kapranov, A.~V.~Zelevinsky, 
{\em Adv. Math.} \textbf{84} (1990) 255-271.


\end{thebibliography}
\end{document}